# Quantum tomography of an entangled three-spin state in silicon

Kenta Takeda,[1,*] Akito Noiri, [1] Takashi Nakajima, [1] Jun Yoneda, [1,†], Takashi Kobayashi, [1] and Seigo Tarucha[1,*]

[1]Center for Emergent Matter Science (CEMS), RIKEN, Wako-shi, Saitama, 351-0198, Japan

[*]Correspondence to: Kenta Takeda (kenta.takeda@riken.jp) or Seigo Tarucha (tarucha@riken.jp)

[†]Present address: School of Electrical Engineering and Telecommunications, University of New South Wales, Sydney, New South Wales 2052, Australia

Abstract:
Quantum entanglement is a fundamental property of coherent quantum states and an essential resource for quantum computing. While two-qubit entanglement has been demonstrated for spins in silicon, creation of multipartite entanglement, a first step toward implementing quantum error correction, has remained challenging due to the difficulties in controlling a multi-qubit array, such as device disorder, magnetic and electrical noises and exacting exchange controls. Here, we show operation of a fully functional three-qubit array in silicon and generation of a three-qubit Greenberger–Horne–Zeilinger (GHZ) state. We obtain a state fidelity of 88.0 percent by quantum state tomography, which witnesses a genuine GHZ-class quantum entanglement that is not biseparable. Our result shows the potential of silicon-based qubit platform for demonstrations of multiqubit quantum algorithms.

Solid-state qubits defined by standard lithography techniques provide a promising route for scaling-up quantum computers [1]. Among their possible candidates, the silicon-based implementations offer a great potential for high-density integration using standard semiconductor manufacturing techniques [2,3]. For the single electron spins in silicon, key ingredients such as single-qubit control fidelities exceeding 99.9 % [4,5], a two-qubit control fidelity of about 98 % [6], quantum non-demolition measurement [7,8], and high-temperature operation [9,10] have been demonstrated. Realization of a large-scale system will require dealing with the error accumulation by quantum error correction [11], in which one qubit information is distributed to a multipartite entangled state. The multipartite entanglement has been demonstrated, for example, in superconducting circuits [12,13], trapped ions [14], and nitrogen-vacancy centers in diamond [15]. Here we demonstrate the simplest example of genuinely multipartite entanglement, a three-qubit GHZ state [16], using a device fabricated in a low-disorder silicon/silicon-germanium (Si/SiGe) substrate. The device enables formation of a controllable triple quantum dot array with a low level of charge noise. We implement individual spin measurements, high-fidelity single-qubit gates, and two-qubit operations carefully designed to mitigate the effects of the low-frequency noise. Finally, we properly combine them to perform a GHZ state preparation protocol and characterize the generated state by quantum state tomography.

Our three-qubit device is a series coupled triple quantum dot fabricated on an isotopically natural Si/SiGe heterostructure wafer (Fig. 1(a)). Nanofabricated overlapping aluminum gates [17] are used to control the triple quantum dot confinement potential (Fig. 1(a)). Each quantum dot is configured to host one electron (Fig. 1(c)) and its spin-up and -down states are used to encode a spin qubit (Fig. 1(b), $Q_1$ to $Q_3$ from left to right). An in-plane external magnetic field of $B_{\mathrm{ext}} = 0.5275\ \mathrm{T}$ is applied to create a Zeeman splitting (~ 18 GHz) much larger than the thermal broadening (~ 40 mK). This enables initialization and measurement of the spin qubits ($Q_1$ and $Q_3$) with neighboring electron reservoirs via energy-selective tunneling [18] - see Fig. 1(c) for bias configuration. The middle spin qubit ($Q_2$), which does not directly couple to any reservoir, can be initialized and measured by the combinations of a resonant SWAP gate [19,20] and the energy-selective tunneling (see Fig. S1 for details). With this initialization and measurement procedure, we obtain high initialization fidelities ($> 99\ \%$ for all qubits) and the readout fidelities of $F_1^{\mathrm{M}} = 87.9\ \%$ ($Q_1$), $F_2^{\mathrm{M}} = 82.9\ \%$ ($Q_2$), and $F_3^{\mathrm{M}} = 89.9\ \%$ ($Q_3$).

The single-qubit control is performed by electric-dipole spin resonance (EDSR). A magnetic field gradient created by a cobalt micro-magnet placed on top of the quantum dot array enables fast and addressable EDSR [21]. In Fig. 2, (a) to (c), the Rabi-chevron patterns observed for the three spin qubits. The resonance frequencies of the three qubits are found to be separated by $\delta E_{12} \approx 435.4\ \mathrm{MHz}$ (between $Q_1$ and $Q_2$) and $\delta E_{23} \approx 523.2$ MHz (between $Q_2$ and $Q_3$) due to the micro-magnet field gradient [7,20,22–24]. These large separations ensure that our single-qubit drive (with $f_{\mathrm{Rabi}} = 6$ MHz) does not rotate the other spins. We then characterize the $T_1$ relaxation times, $T_2^*$ inhomogeneous dephasing times, and $T_2^{\mathrm{echo}}$ Hahn echo times (Fig. S2). We find $T_1 = 4.30$ ms ($Q_1$), 2.67 ms ($Q_2$), and 1.31 ms ($Q_3$), which are long enough to perform single-shot spin readout, but shorter than what are typically expected for electron spins in silicon probably due to spin-valley mixing [25]. The dephasing times of $T_2^* = 1.82$ μs ($Q_1$), 1.69 μs ($Q_2$), and 1.69 μs ($Q_3$) are similar to those reported for natural silicon spin-1/2 qubits [22–24] and are most likely limited by the fluctuation of 4.7% $^{29}$Si nuclear spins in natural silicon. Hahn echo extends the dephasing times to $T_2^{\mathrm{echo}} = 28.1$ μs ($Q_1$) 20.5 μs ($Q_2$), and 45.8 μs ($Q_3$). The average single-qubit control fidelities are examined through the Clifford-based randomized benchmarking [26] and found to be 99.43 % ($Q_1$), 99.57 % ($Q_2$), and 99.91 % ($Q_3$) (Fig. 2, (d) to (f)). To perform ideal single-qubit control for multi-qubit input states, the exchange interactions between the spins should be low compared to $f_{\mathrm{Rabi}}$. However, in the current experiment, the residual exchange is relatively large between $Q_2$ and $Q_3$, resulting in a reduction of the control fidelities of $Q_2$ and $Q_3$ by 0.6 % (see Supplementary Information). Nevertheless, these fidelities are high enough to perform the quantum state tomography measurements. To create the quantum entanglement, we utilize controlled phase (CZ) gates between the neighboring spin qubits. The exchange coupling $J_{ij}$ between the neighboring spin qubits $Q_i$ and $Q_j$ acts as an effective Ising type interaction $(hJ'_{ij}/4)\sigma_z^i \otimes \sigma_z^j$ under the large micro-magnet local Zeeman gradient $\delta E_{ij}$, where $J'_{ij} = \sqrt{J_{ij}^2 + (\delta E_{ij})^2} - \delta E_{ij}$ and $h$ is the Planck's constant [27]. The time evolution under this interaction for a period of $(2J'_{ij})^{-1}$ is equivalent to a CZ gate between $Q_i$ and $Q_j$ ($CZ_{ij}$) up to local single-qubit phases. To control $J'_{12}$ ($J'_{23}$) in time-domain, we utilize fast gate voltage pulses applied on the B2 (B3) gate (Fig. 1(b)). When turned on, $J'_{12}$ and $J'_{23}$ are nominally 2.8 MHz and 12.5 MHz, respectively (Fig. S3, (a) to (d)). Additional linear compensations are applied to the plunger gates P1, P2, and P3 in order to keep the triple dot charge configuration near the symmetric operation point (see Supplementary Information). At this point, the charge-noise-induced dephasing is minimized since $J'_{ij}$ becomes first-order insensitive to detuning fluctuations [28,29].

We combine the single-qubit and CZ gates to perform a three-qubit entangling operation. While three qubits can be entangled simply by sequentially applying CZ gates between neighboring qubits ($CZ_{12}$ and $CZ_{23}$) as demonstrated in other systems [12,13,30], the qubits are vulnerable to the low-frequency noise during manipulation. To fully leverage the long intrinsic phase coherence times of spins, here we implement a decoupled three-qubit entangling operation by extending a decoupled CZ gate developed in ref. [23] to a three qubit system (Fig. 2(g)). In this sequence, the CZ gates are separated into $\sqrt{\mathrm{CZ}}$ gates with π pulses inserted in the middle. As in the standard Hahn echo experiments, the $\pi$ pulses reverse the non-conditional phase accumulation during free evolution, and therefore decouple the quasi-static single-qubit phase noise (for example, the low-frequency nuclear magnetic and charge noises and the local phase accumulated by the CZ pulses). The action of our three-qubit operation is measured by a quantum gate sequence shown in Fig. 2(g). Here, $Q_2$ is used as a control qubit and the phase of the second $\pi/2$ pulse ($\phi$) is varied to detect the phase accumulation on $Q_1$ and $Q_3$. Figure 2(h) (2(i)) shows the spin-up probabilities of $Q_1$ ($Q_3$) measured as a function of $\phi$. Additional Z rotations at the end are required to adjust the additional phases originating from the $\pi$ pulses. The exchange pulse durations are tuned up so that the conditional phase accumulation is $\approx \pi$. The phases are typically calibrated within $\pm 0.01\pi$ from the target values. As a benchmark of our CZ gate quality, we measure the fidelities of two-qubit Bell states using $Q_2$ and $Q_3$. We obtain an average Bell state fidelity of 94.1 % and concurrence of 0.929 (see Fig. S4), which compare favorably to the values previously reported in other silicon quantum-dot-based qubit devices (fidelities of 78 to 89 % and concurrences of 0.73 to 0.82 in Refs. [6,22,23]). The result indicates that our CZ gate fidelity is reasonably high, although further assessment using two-qubit Clifford-based randomized benchmarking [6] is necessary to extract the CZ gate fidelity and it remains for future study.

Now we turn to the generation of three-qubit entanglement. Figure 3(a) shows our quantum gate sequence to generate a three-qubit GHZ state $|\mathrm{GHZ}\rangle = (|\uparrow\uparrow\uparrow\rangle + |\downarrow\downarrow\downarrow\rangle)/\sqrt{2}$. The sequence is similar to the one in Fig. 2(g), but with the first rotation on $Q_2$ replaced with a $\mathrm{Y}/2$ gate and some phase corrections. After the state preparation, we apply a single-qubit pre-rotation $(\mathrm{I}, \mathrm{X}/2, \mathrm{Y}/2)$ on each qubit in order to rotate the measurement axis. For each of the 27 combinations of pre-rotations, we average 2,000 single-shot readout outcomes to obtain the eight probabilities projected to the computational basis. The measurement errors are removed by correcting the obtained probabilities based on the spin-up and -down readout fidelities (see Supplementary Information). We reconstruct a density matrix $\rho$ using maximum likelihood estimation so that $\rho$ is physical, i.e. Hermitian, positive-semidefinite, and unit trace (see Supplementary Information). Figure 3(b) shows the real part of the measured density matrix $\mathrm{Re}(\rho)$ in the computational basis (see Fig. S5 for the imaginary part). As expected for a GHZ state, there are four peaks at the corners. Figure 3(c) shows the expectation values for non-trivial 63 Pauli operators, which are also in a good agreement with the ideal GHZ state shown as the open black boxes. By comparing the measured state with the ideal one, we obtain a state fidelity of $F_{\mathrm{GHZ}} = \langle\mathrm{GHZ}|\rho|\mathrm{GHZ}\rangle = 0.880 \pm 0.007$, which is comparable to the value obtained in the first demonstration in a superconducting transmon three-qubit device [12] and high enough to test quantum algorithms [31,32].

To understand the properties of the generated state, we evaluate entanglement witness operators. The measured state fidelity is useful to distinguish two types of maximally entangled three-qubit states, the GHZ-class and W-class states. Only the GHZ-class state becomes completely separable after loss and/or measurement of one qubit information, and therefore is a useful kind for quantum error correction. Our measured state strictly belongs to the GHZ-class because any W-class state $\rho_{\mathrm{W}}$ satisfies $\langle \mathrm{GHZ}|\rho_{\mathrm{W}}|\mathrm{GHZ}\rangle \leq 0.75$ . To further infer the degree of the generated three-qubit entanglement, we evaluate a witness function $M = \langle \mathrm{XXX}\rangle - \langle \mathrm{XYY}\rangle - \langle \mathrm{YXY}\rangle - \langle \mathrm{YYX}\rangle$, which satisfies the Mermin-Bell inequality $|M| \leq 2$ for any biseparable states [33]. We measure $M = 3.47 \pm 0.05$, which violates the inequality by more than 28 standard deviations. The violation is not free from the loopholes in locality and detection due to the limited readout fidelities, but it is nonetheless an important indication of three-qubit entanglement.

We believe that there is still a room for enhancement of the state fidelity by the efforts to improve the quantum gates. The obtained state fidelity is comparably limited by the eight single-qubit and four $\sqrt{\mathrm{CZ}}$ gates in the GHZ state generation protocol in Fig. 3(a). While the single-qubit gates are expected to have higher fidelities (>~ 99%) than the CZ gates, they will have a non-negligible to the measured state infidelity of 12 %. The single-qubit control fidelity can be improved by using an isotopically enriched $^{28}$Si/SiGe material to reduce the magnetic noise [4]. In contrast, improvement of the two-qubit fidelity will require reduction of charge noise, which may be possible with optimization of the device structure [34].

In conclusion, we show operation of a three-qubit device in silicon and have performed generation and measurement of a three-qubit GHZ state. The generated state, fully characterized by quantum state tomography, shows a high fidelity and properties of genuine three-qubit entanglement. We anticipate that our results will enable exploration of multi-spin correlations and demonstration of multi-qubit algorithms such as quantum error correction in scalable silicon-based quantum computing devices.

We thank the Microwave Research Group in Caltech for technical support. This work was supported financially by Core Research for Evolutional Science and Technology (CREST), Japan Science and Technology Agency (JST) (JPMJCR15N2 and JPMJCR1675), MEXT Quantum Leap Flagship Program (MEXT Q-LEAP) grant Nos. JPMXS0118069228, and JSPS KAKENHI grant Nos. 26220710, 16H02204, 17K14078, 18H01819, 19K14640, and 20H00237. T.N. acknowledges support from The Murata Science Foundation Research Grant.

Figures and tables:

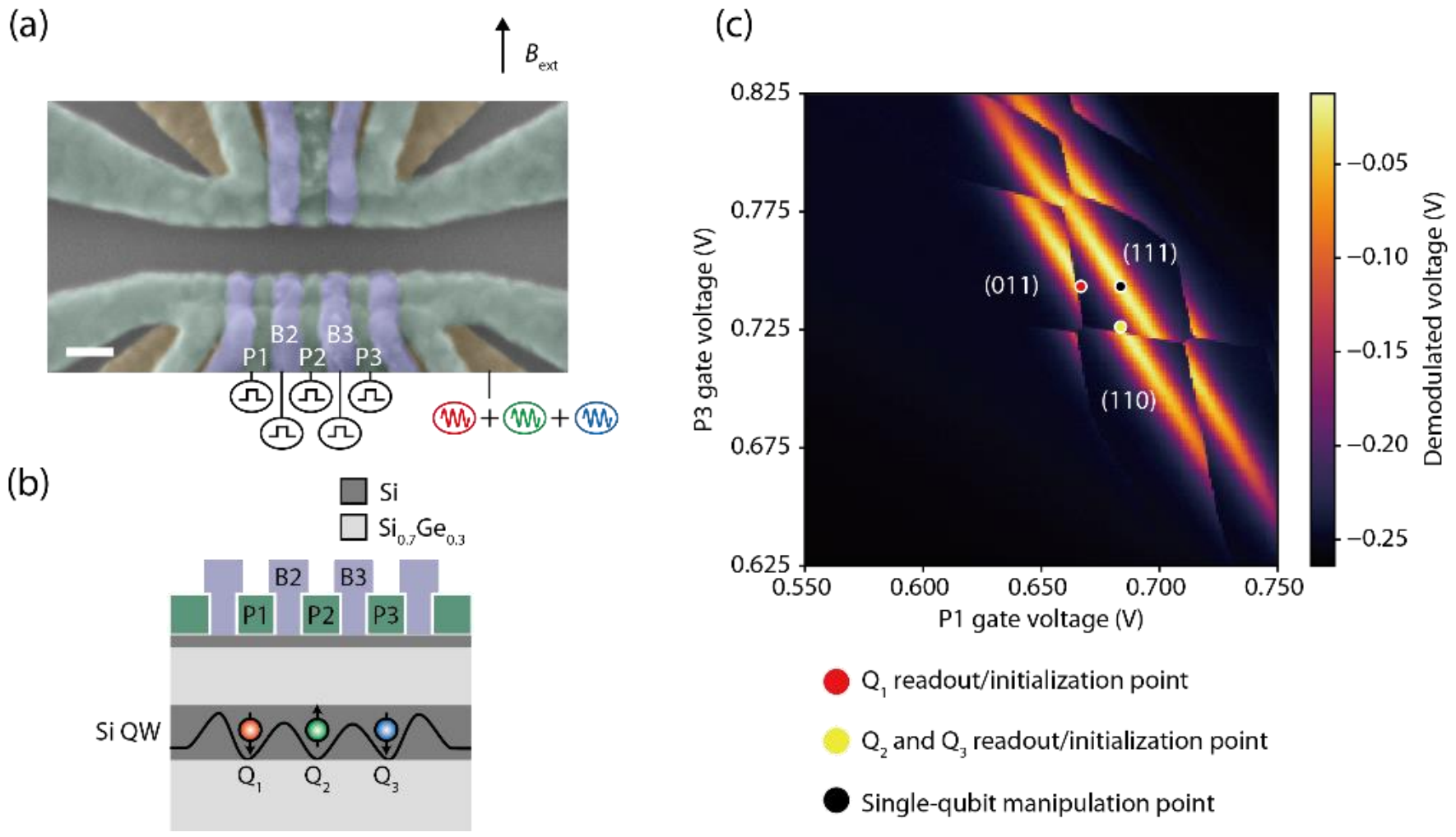

**Fig. 1. Device and experimental setup.** (a) False colored scanning electron microscope image of the device. The scale bar is 100 nm. Three layers of overlapping aluminum gates are fabricated on top of a Si/SiGe heterostructure wafer. The plunger (green) and barrier (purple) gates are used to control the confinement potential and accumulate the reservoirs. The screening gates (ocher) restrict the electric field of the plunger and barrier gates. The lower channel is used as a triple quantum dot array and the upper channel is used as a charge sensor. Gates P1, P2, P3, B2, and B3 are connected to high-bandwidth coaxial cables to apply fast voltage pulses. (b) Cross-sectional schematic of the device. The black solid line in the silicon quantum well (Si QW) illustrates the triple quantum dot confinement potential. (c) Charge stability diagram with the exchange interactions turned off. The color plot shows the demodulated voltage of the radio-frequency sensor quantum dot as a function of the voltages applied to P1 and P3 gates. Kinks appear when the charge occupation of the triple quantum dot ($n_1 n_2 n_3$) changes, where $n_1$, $n_2$ and $n_3$ are the respective numbers of electrons in the left, middle and right quantum dots. The background signal variation is due to the sensor dot Coulomb oscillation.

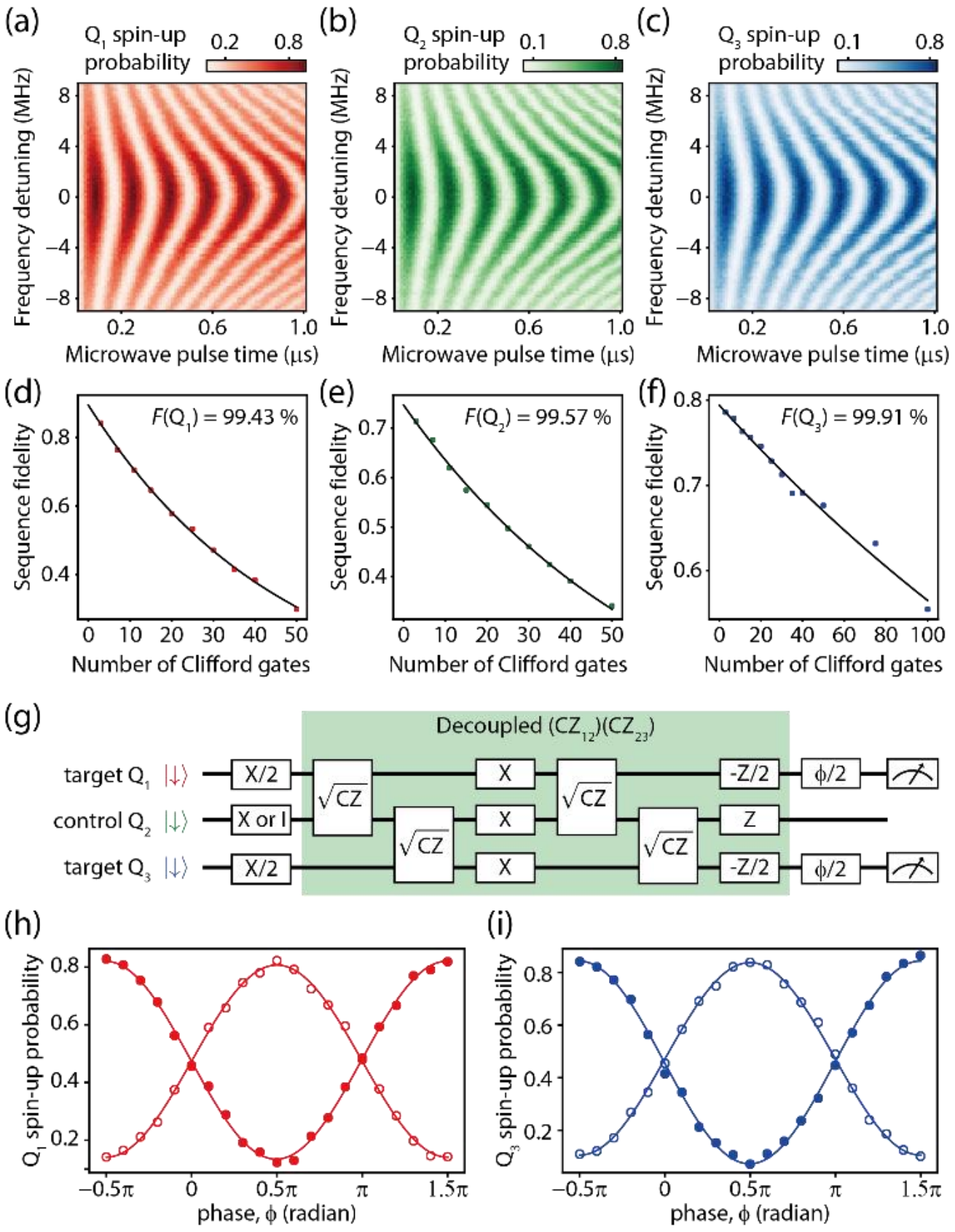

**Fig. 2. Single-qubit and controlled phase operations.** (a to c) Single-qubit Rabi chevron patterns of each qubit. Each spin state is read out right after the manipulation stage without sequential readout for the three spins. The frequency offsets are 17789.15 MHz ($Q_1$), 18224.5 MHz ($Q_2$), and 18747.7 MHz ($Q_3$). (d to f) Single-qubit randomized benchmarking results for each qubit. The visibilities are 0.894, 0.746, and 0.794 for $Q_1$, $Q_2$, and $Q_3$. (g) Quantum gate sequence used to tune-up the CZ gates targeting both qubits $Q_1$ and $Q_3$. A similar gate sequence, but with the target and control qubits swapped, is used to tune-up the unconditional phase accumulated on $Q_2$ during the exchange pulses. $\phi/2$ indicates a $\pi/2$ rotation around an axis in the xy plane - when $\phi = 0$ $(0.5\pi)$, it is around the x (y) axis. Here $\sqrt{\mathrm{CZ}}$ is defined as $\mathrm{diag}(1,1,1,i)$. (h and i) Measured $Q_1$ and $Q_3$ spin-up probabilities for two different control bit states of $Q_2$. The filled (open) circles represent the result when $Q_2$ is spin-down (up). The solid lines are the sinusoidal fitting curves. From the phase offsets of the sinusoids, we obtain the conditional phase shifts of $(1.004 \pm 0.005)\pi$ radian for $Q_1$ and $(1.003 \pm 0.004)\pi$ radian for $Q_3$. We note that we also apply a small $(< 0.1\pi)$ phase correction to each qubit in order to account for the imperfect cancellation of the non-conditional phase accumulation during the echo sequence.

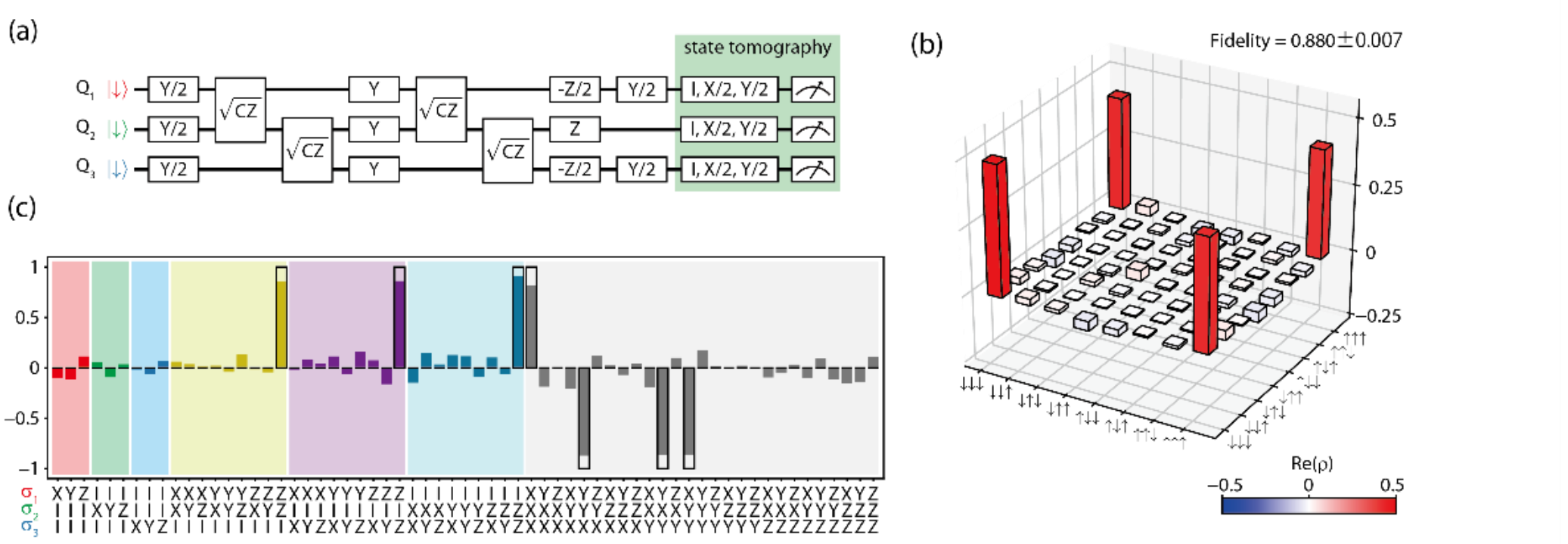

**Fig. 3. Three-qubit entanglement generation and measurement.** (a) Quantum gate sequence used for GHZ state generation and measurement. It is similar to the double CZ sequence in Fig. 2(g). The key difference is that the first control pulse for $Q_2$ is replaced with a Y/2 pulse to prepare a superposition state. In addition, the X pulses are replaced with the Y pulses to obtain a final state with desired phase components. (b) Real part of the measured density matrix in the computational basis. The imaginary part can be found in Supplementary information. (c) Measured expectation values for Pauli operators. An obvious expectation value $\langle III \rangle = 1$ is omitted in the plot. The open boxes represent the expectation values for an ideal three-qubit GHZ state.

## Supplementary Information for "Quantum tomography of an entangled three-spin state in silicon"

Kenta Takeda,[1,*] Akito Noiri,[1] Takashi Nakajima,[1] Jun Yoneda,[1,†], Takashi Kobayashi,[1] and Seigo Tarucha[1,*]

[1]Center for Emergent Matter Science (CEMS), RIKEN, Wako-shi, Saitama, 351-0198, Japan

[†]Present address: School of Electrical Engineering and Telecommunications, University of New South Wales, Sydney, New South Wales 2052, Australia

Device

The triple quantum dot device is fabricated using an isotopically natural, undoped Si/SiGe heterostructure with a mobility of 300,000 $\mathrm{cm^2/V\cdot s}$ at an electron density of $n = 5 \times 10^{11}\ \mathrm{cm^{-2}}$ and a temperature of 2 K. Phosphorus ion implantation is used to make the ohmic contacts. Standard electron-beam lithography and lift-off techniques are used to fabricate the overlapping aluminum gates and the micro-magnet. The micro-magnet is a stack of Ti/Co/Al films with thicknesses of 10/250/20 nm. The 20-nm-thick aluminum film is expected to serve as an anti-oxidation layer. The effective low-frequency charge noise measured in this device is $\mathrm{S}(f = 1\ \mathrm{Hz}) = 0.2\ \mu\mathrm{eV}/\sqrt{\mathrm{Hz}}$.

Measurement setup

The sample is cooled down in a dry dilution refrigerator (Oxford Instruments Triton) to a base electron temperature of 40 mK. The dc gate voltages are supplied using a 24-channel digital-to-analog converter (QDevil ApS QDAC). Home-made cryogenic low-pass filters with a cutoff frequency of 160 Hz are used to filter its outputs. The gate voltage pulses are generated by an arbitrary waveform generator (Tektronix AWG5208) and are filtered by Mini Circuits SBLP-39+ Bessel low-pass filters. The EDSR microwave pulses are generated using three I/Q modulated signal generators (two Keysight E8267D and a Keysight E8257D with a Marki microwave MMIQ-0626H I/Q mixer). The waveforms for I/Q modulation are generated by another AWG5208 unit triggered by the AWG used for generating the gate voltage pulses. The microwave signals are sideband-modulated by 20 MHz from the baseband frequencies in order to avoid the unintentional spin rotation due to leakage (the typical isolation is ~ 50 dBc after calibration of the I/Q imbalances and the dc offsets). Unless noted, the microwave pulses have a square envelope with $\sim$10 ns rise and fall times. During the initialization and manipulation stages, additional pulse modulations are used to provide further isolation of the microwave signals. The radio-frequency pulses used for resonant SWAP gates are generated by a pulse-modulated Agilent E4432B signal generator. The radio-frequency pulses are combined to the EDSR microwave pulses and applied to a screening gate.

Charge sensing is performed using a radio-frequency reflectometry technique [35,36]. The charge sensor quantum dot is connected to a tank circuit with a resonance frequency of 181.5 MHz. The reflected signal is amplified by a cryogenic amplifier (Caltech CITLF1) and further amplified and demodulated at room-temperature. The demodulated signal is digitized using an AlazarTech ATS9440 digitizer card at a sampling rate of 5 MSa/s. The digitized signal is filtered by a 7th order digital Bessel low-pass filter with a cutoff-frequency of 0.4 MHz in order to achieve a signal-to-noise ratio high enough for the single-shot spin readout.

## Exchange gates

The exchange interactions are controlled through gate voltage pulses. To keep the quantum dot potential near the symmetric operation point during the exchange pulses, we utilize the virtual gate technique [37]. We measure the gate-voltage-induced shift of charge transition line of every quantum dot to extract the lever arms of the gates P1, P2, P3, B2, and B3. From the measured lever arms, we construct the virtual barrier gates $\delta V_{12}$ and $\delta V_{23}$ as follows:

$$(-0.218, 1, -0.180, 0, 0.005)\delta V_{12} = (\delta V_{\mathrm{P1}}, \delta V_{\mathrm{B2}}, \delta V_{\mathrm{P2}}, \delta V_{\mathrm{B3}}, \delta V_{\mathrm{P3}}),$$
$$(0.018, 0, -0.244, 1, -0.226)\delta V_{23} = (\delta V_{\mathrm{P1}}, \delta V_{\mathrm{B2}}, \delta V_{\mathrm{P2}}, \delta V_{\mathrm{B3}}, \delta V_{\mathrm{P3}}).$$

We typically use the virtual barrier gate voltage pulses $\delta V_{12} = 0.13\ \mathrm{V}$ and $\delta V_{23} = 0.075\ \mathrm{V}$ to turn on $J'_{12} = 2.8\ \mathrm{MHz}$ and $J'_{23} = 12.8\ \mathrm{MHz}$ respectively (Fig. S3, a and b). The residual exchange interactions are measured to be $J_{12}^{\mathrm{off}} < 0.1\ \mathrm{MHz}$ and $J_{23}^{\mathrm{off}} = 1.2\ \mathrm{MHz}$ using Ramsey interferometry (Fig. S3, c and d). $J_{12}^{\mathrm{off}}$ is below our detection limit and has a minor effect on our measurements. $J_{23}^{\mathrm{off}}$ is noticeably larger than our qubit linewidths and affects the single-qubit control fidelities. This effect is mitigated by shifting the qubit drive frequencies of $Q_2$ and $Q_3$ by 0.6 MHz from their resonance frequencies measured with all qubits initialized to spin-down. While the tunnel coupling between the middle and right quantum dots could be quenched by reducing the dc voltage applied on B3 gate, it resulted in the drastic reduction of the $T_1$ relaxation time of $Q_3$. The valley splitting of the right quantum dot most likely changes with the barrier gate voltage [38] and it enhances spin-valley hotspot relaxation at the magnetic field used. Although the external magnetic field could be tuned to avoid the spin-valley relaxation, we are not able to take this approach here due to the limited microwave frequency range available in our measurement setup.

## State tomography

In order to remove the state preparation and measurement errors, we characterize the initialization and readout fidelities. First, we measure the initialization fidelity of each qubit using a similar scheme as in Ref. [23]. As we do not see any enhancement of the initialization fidelities even after waiting 50 msec (longer than $9T_1$) at the single-qubit operation point, we conclude that the initialization fidelities are high (> 99 %) for all qubits. Since these are much higher than the readout fidelities, throughout this work, the initialization infidelities are ignored. The spin-down (up) readout fidelity $F_{\downarrow i}(F_{\uparrow i})$ is determined from the measured spin-up probability after the standard initialization protocol followed by an I (X) gate. The typical readout fidelities are measured to be $F_{\downarrow 1} = 0.906$, $F_{\uparrow 1} = 0.852$, $F_{\downarrow 2} = 0.948$, $F_{\uparrow 2} = 0.711$, $F_{\downarrow 3} = 0.955$, and $F_{\uparrow 3} = 0.844$. The readout fidelities in the main text are defined as $F_i = (F_{\downarrow i} + F_{\uparrow i})/2$. The spin-up readout fidelities of $Q_1$ and $Q_2$ are largely limited by the spin relaxation during the sequential readout. Using these parameters, we correct the measured probabilities $\boldsymbol{P}_{\mathrm{M}} = (p_{\downarrow\downarrow\downarrow}, \dots, p_{\uparrow\uparrow\uparrow})$ as $\boldsymbol{P} = (F_1 \otimes F_2 \otimes F_3)^{-1} \boldsymbol{P}_{\mathrm{M}}$, where $F_i = \begin{pmatrix} F_{\downarrow i} & 1 - F_{\uparrow i} \\ 1 - F_{\downarrow i} & F_{\uparrow i} \end{pmatrix}$ and $\boldsymbol{P}$ is the probabilities used for the input of maximum likelihood estimation.

Maximum likelihood estimation is performed to restrict the density matrix to be physical. A physical density matrix can be written as $\rho = T^{\dagger}T/\mathrm{Tr}(T^{\dagger}T)$ where $T$ is a complex lower triangular matrix with real diagonal elements. $T$ has 64 real parameters $\boldsymbol{t} = (t_1, \dots, t_{64})$ and we minimize the cost function

$$C(\boldsymbol{t}) = \sum_{\nu=1}^{64} \frac{(\langle \psi_\nu | \rho(\boldsymbol{t}) | \psi_\nu \rangle - p_\nu)^2}{2\langle \psi_\nu | \rho(\boldsymbol{t}) | \psi_\nu \rangle},$$

where $p_\nu$ is the measured probability projected at $|\psi_\nu\rangle$. To determine the 64 parameters, the projection outcomes for linearly independent pre-rotations $(\mathrm{I}, \mathrm{X}/2, \mathrm{Y}/2, \mathrm{X})^{\otimes 3}$ are used. To remove the error that could be introduced by the X pre-rotation, the projection outcomes including the X pre-rotations are calculated from the corresponding I rotation outcomes [23]. The errors of the maximum likelihood estimation results are obtained by Monte-Carlo method assuming that the measured single-shot probabilities follow multinomial distributions [6,23]. We fit each of the resulting distributions with a Gaussian function to extract its standard deviation.

Two-qubit Bell state tomography

As a benchmark of our two-qubit CZ gate, we perform Bell state tomography on $Q_2$ and $Q_3$. The experiment is a reduced version of the 3-qubit GHZ state tomography. The readout errors are removed using the measured readout fidelities and maximum likelihood estimation is used to reconstruct the density matrices. The quantum gate sequence for the Bell state tomography is shown in Fig. S4(a). By applying appropriate Z gates after the second $\sqrt{\mathrm{CZ}}$ gate, we can create four Bell states $\Phi^+ = (|\uparrow\uparrow\rangle + |\downarrow\downarrow\rangle)/\sqrt{2}$, $\Phi^- = (|\uparrow\uparrow\rangle - |\downarrow\downarrow\rangle)/\sqrt{2}$, $\Psi^+ = (|\downarrow\uparrow\rangle + |\downarrow\uparrow\rangle)/\sqrt{2}$, and $\Psi^- = (|\downarrow\uparrow\rangle - |\uparrow\downarrow\rangle)/\sqrt{2}$ (Fig. S4, b to e). We obtain the state fidelities relative to the target states of 0.942 ($\Phi^+$), 0.933 ($\Phi^-$), 0.950 ($\Psi^+$), and 0.940 ($\Psi^-$), and the concurrences of 0.950 ($\Phi^+$),0.906 ($\Phi^-$),0.923 ($\Psi^+$), and 0.935 ($\Psi^-$).

Imaginary part of the experimental GHZ state

Fig. S5 shows the imaginary part of the experimentally obtained GHZ state in Fig. 3 in the main text.

Error analysis

Except for the ones for the state tomography results, all errors represent the estimated standard errors for the best-fit values.

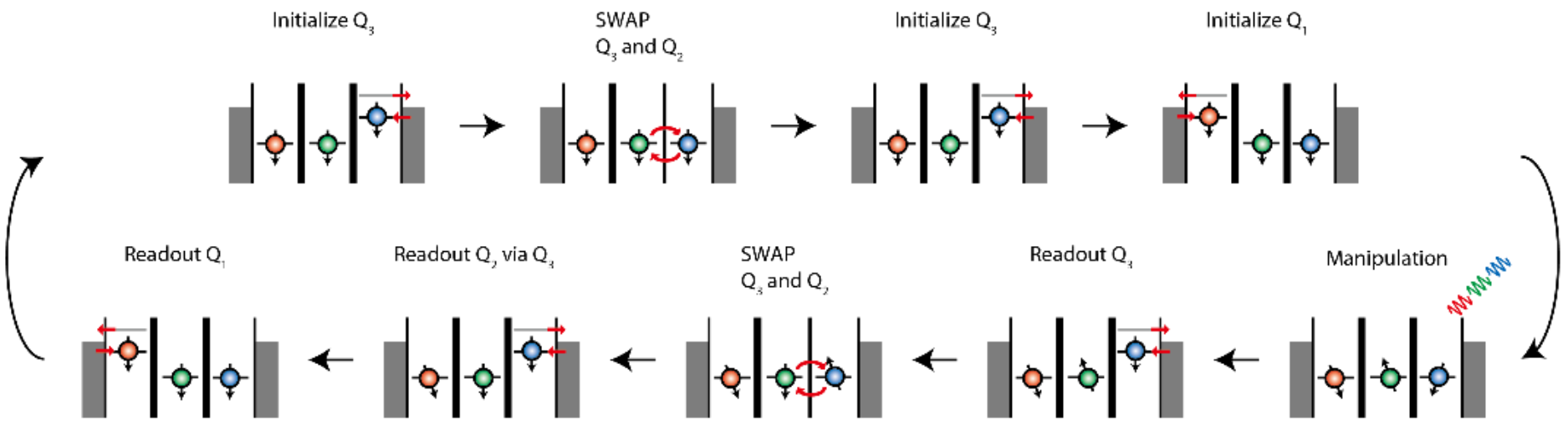

**Fig. S1. Initialization and measurement protocol.** Initialization and readout procedure. The spin readout and initialization for $Q_1$ is performed near the (111)-(011) boundary, whereas for $Q_2$ and $Q_3$ it is performed near the (111)-(110) boundary. Note that $Q_2$ cannot be directly read out through the reservoir as the co-tunneling rate between the (111) and (101) states is too small compared to the spin relaxation rate $T_1^{-1}$. The dwell times are 450 μs for $Q_3$ initialization, 750 μs for $Q_1$ initialization, 300 μs for $Q_3$ readout, and 750 μs for $Q_1$ readout. The resonant SWAP pulses are 0.25-μs-long and it corresponds to an exchange Rabi frequency of 2 MHz. The resonance frequency is typically around 410 MHz. The initialization stage may be redundant because all the three spins are ideally initialized to spin-down after the readout stage, but it is employed to increase the initialization fidelities.

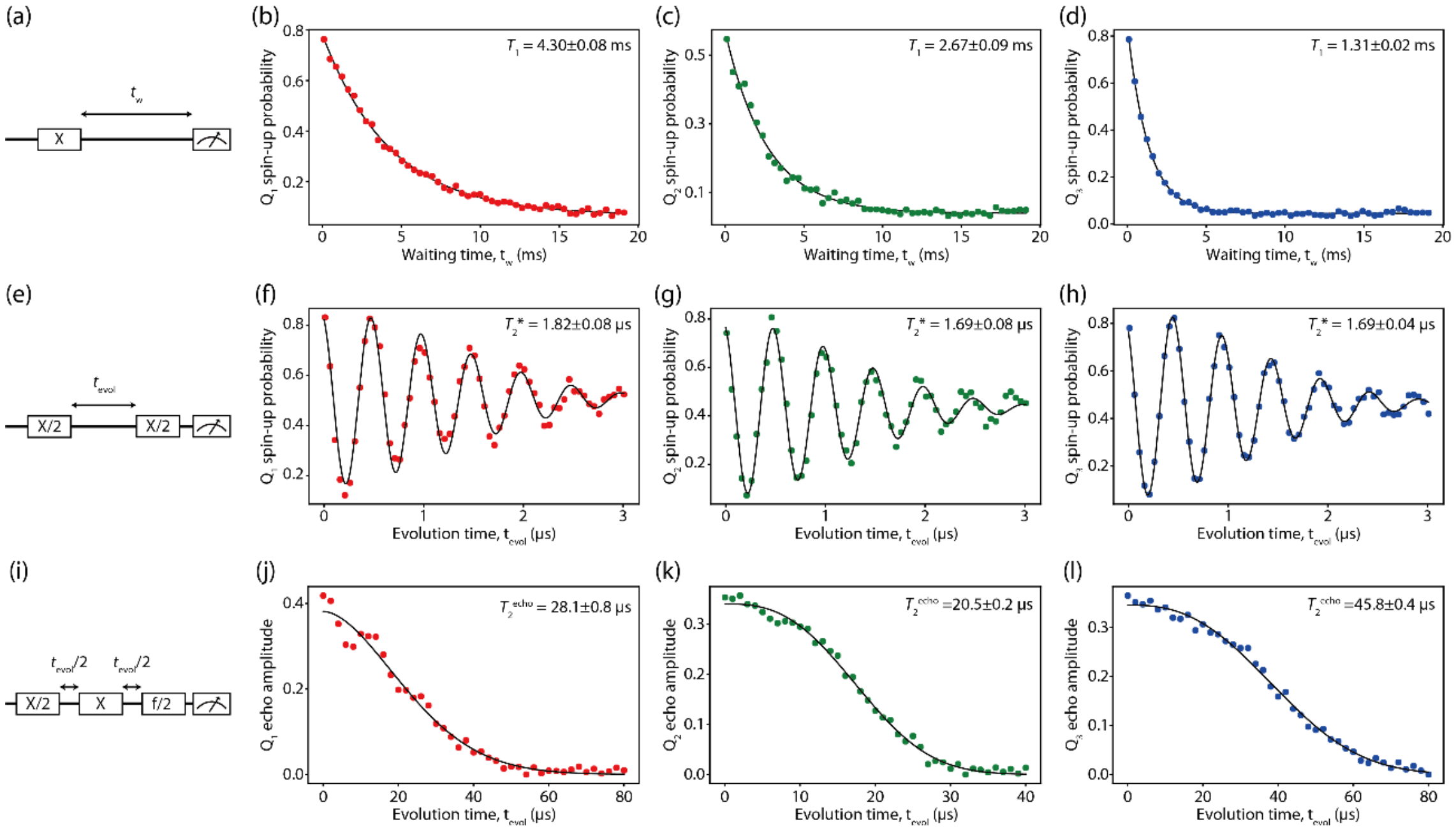

**Fig. S2. Single-qubit characterization.** All qubits are initialized to spin-down before the manipulation stage and only one of the qubits is read out right after the manipulation stage unless noted. The exchange interactions are turned off. (a to d) $T_1$ measurements. First a spin-up state is prepared using an X pulse. Then we vary the waiting time of $t_{\mathrm{w}}$ at the single-qubit manipulation point before performing single-shot measurement. In this $T_1$ measurement, all three spins are sequentially read out and therefore the visibilities of $Q_1$ and $Q_2$ are decreased by $T_1$ relaxation during the readout stage. Note that the visibility of $Q_3$ is unaffected by the sequential readout. (e to h) Ramsey interferometry measurements. First a $\pi/2$ pulse (+2 MHz detuned from the resonance frequency) is applied to rotate the spin state to the xy-plane in the Bloch sphere. After an evolution time of $t_{\mathrm{evol}}$, another $\pi/2$ pulse is applied to project the spin state in the z-axis for measurement. The black solid curves are the fit with Gaussian decay. The integration time is 75.8 sec for all qubits. (i to l) Hahn echo measurements. Each fitting curve is given by $p_{\uparrow}(t_{\mathrm{evol}}) = A\exp(-(t_{\mathrm{evol}}/T_2^{\mathrm{echo}})^{\gamma}) + B$, where $A$ and $B$ are the constants to account for the readout fidelities and $\gamma$ is an exponent. The exponents are found to be $\gamma = 1.79 \pm 0.12$ ($Q_1$), $2.75 \pm 0.10$ ($Q_2$), and $2.61 \pm 0.09$ ($Q_3$).

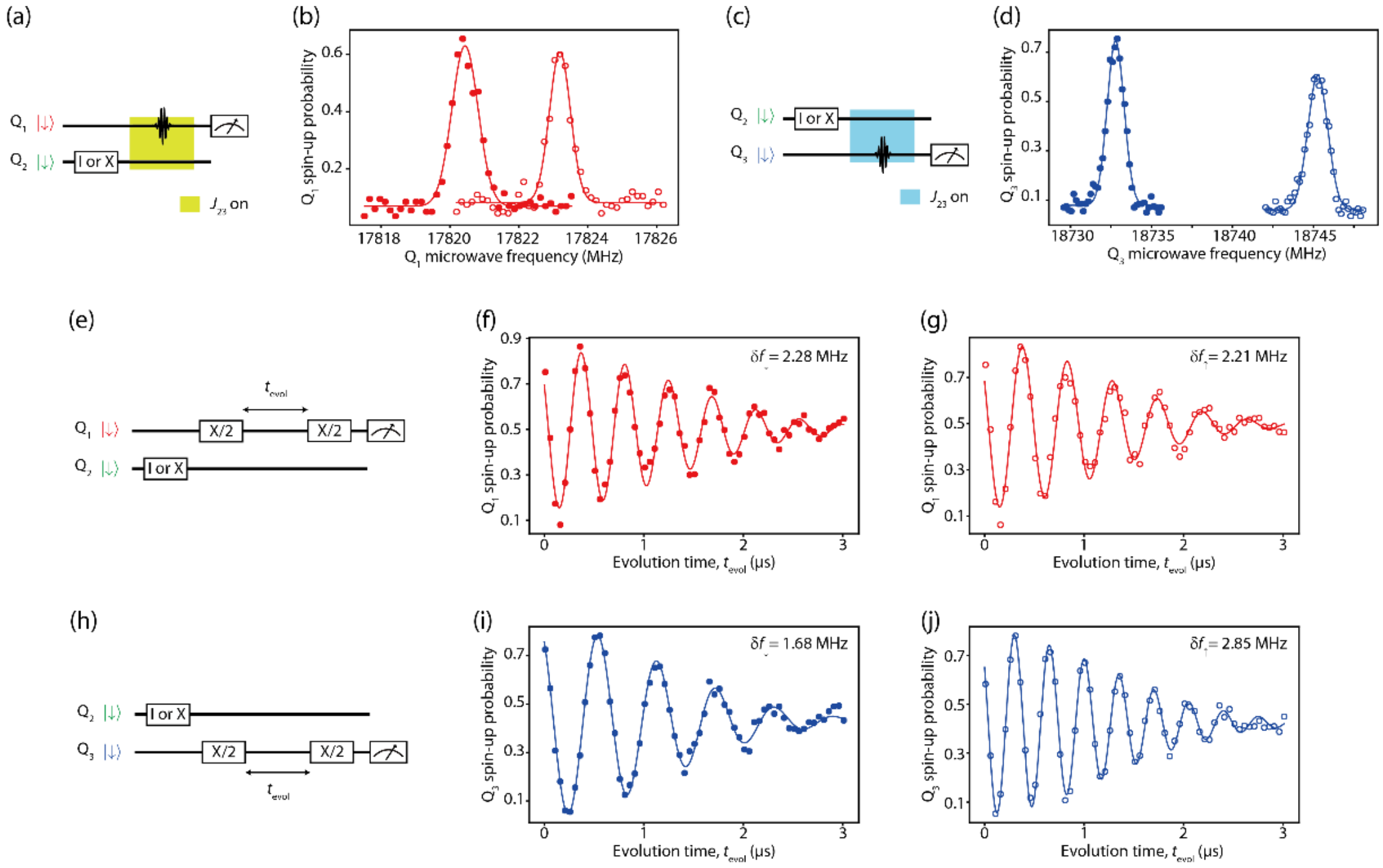

**Fig. S3. Measurements of exchange interactions.** (a and b) Controlled-rotation for $Q_1$ and $Q_2$. The measurement is performed to probe $J_{\mathrm{on}}^{12}$. First, $Q_1$ and $Q_2$ are initialized to spin-down. To prepare a spin-up control qubit ($Q_2$) state, an X pulse is applied. After tuning on $J_{12}$ by a gate voltage pulse, a low-power Gaussian microwave pulse (truncated at $\pm 2\sigma$) is applied to induce a controlled-rotation. The filled (open) circles show the measured spin-up probabilities with the control qubit spin-down (up). The solid lines are Gaussian fitting curves. From the separation of the two peaks, we obtain $J_{12} = 2.75 \pm 0.02$ MHz. (c and d) Similar controlled rotation measurement for $Q_2$ and $Q_3$. We obtain $J_{23} = 12.50 \pm 0.02$ MHz from this measurement. (e) Ramsey experiment to extract $J_{12}^{\mathrm{off}}$. We perform two Ramsey measurements of $Q_1$ with different control qubit ($Q_2$) states. The difference of qubit frequency detuning is equivalent to $J_{12}^{\mathrm{off}}$. (f) Ramsey measurement result when $Q_2$ is spin-down. The red circles are the measured $Q_1$ spin-up probabilities and the black solid curve shows a fit with Gaussian decay. From the oscillation frequency of the decay curve, we extract $\delta f_{\downarrow} = 2.28 \pm 0.01$ MHz. (g) Measurement similar to the one in (f) when $Q_2$ is spin-up. We extract $\delta f_{\uparrow} = 2.21 \pm 0.01$ MHz. Since the difference between $\delta f_{\downarrow}$ and $\delta f_{\uparrow}$ is below the stochastic fluctuation of the frequency detuning, we conclude that $J_{12}^{\mathrm{off}}$ is below our detection limit. Note that each frequency error shows one standard deviation of the fitting parameter. (h to j) Ramsey experiments to extract $J_{23}^{\mathrm{off}}$. We obtain $J_{23}^{\mathrm{off}} = \delta f_{\uparrow} - \delta f_{\downarrow} = 1.17 \pm 0.01$ MHz from these measurements.

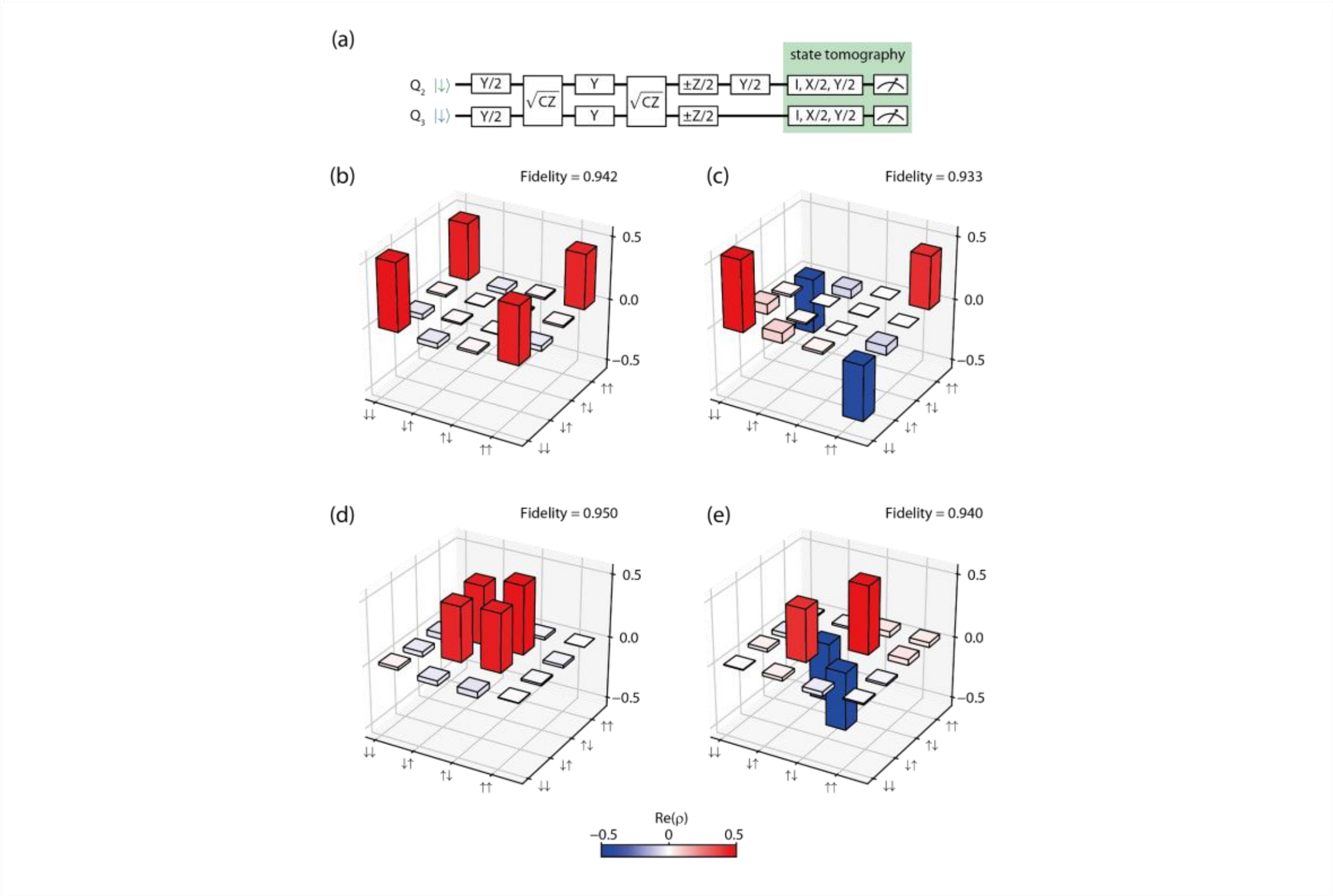

**Fig. S4. Bell state tomography using $Q_2$ and $Q_3$.** (a) Quantum gate sequence for Bell state creation and state tomography. By modifying the phase gates after the second $\sqrt{\mathrm{CZ}}$ pulse, we can create all four Bell states. (b to e), Real parts of the measured density matrices for four Bell states, $\Phi^+ = (|\uparrow\uparrow\rangle + |\downarrow\downarrow\rangle)/\sqrt{2}$ (B), $\Phi^- = (|\uparrow\uparrow\rangle - |\downarrow\downarrow\rangle)/\sqrt{2}$ (C), $\Psi^+ = (|\downarrow\uparrow\rangle + |\downarrow\uparrow\rangle)/\sqrt{2}$ (D), and $\Psi^- = (|\downarrow\uparrow\rangle - |\uparrow\downarrow\rangle)/\sqrt{2}$ (E) .

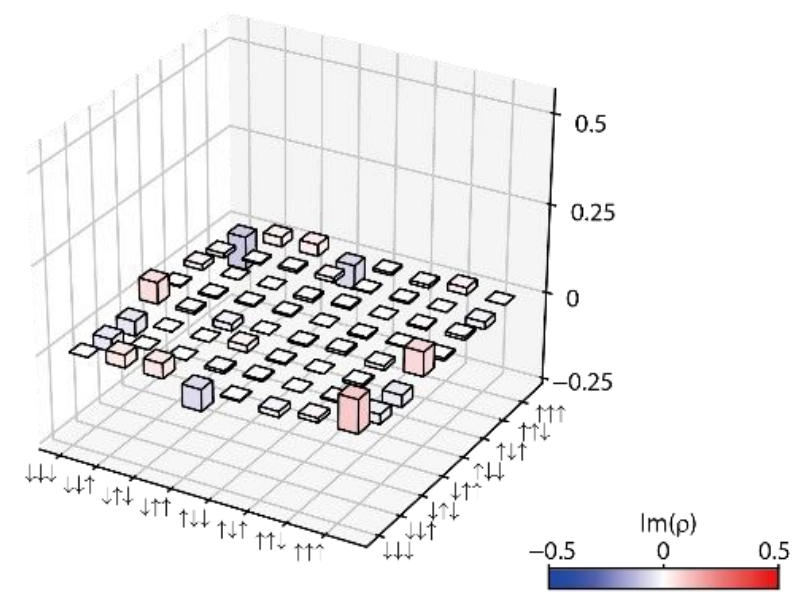

**Fig. S5. Imaginary part of the experimental GHZ state.** The imaginary part is all zero for an ideal GHZ state. Here, the maximum absolute value of the matrix elements is 0.09.